\documentclass[fleqn,onecolumn,11pt]{wlscirep}
\usepackage{bm}
\usepackage{color}
\usepackage{color}
\usepackage{alltt,dsfont}
\usepackage{appendix}
\usepackage{amsmath,amssymb}
\usepackage{braket}
\usepackage[capitalise]{cleveref}
\usepackage{ulem}
\usepackage{float}
\usepackage{graphicx,subfigure,bbm}

\usepackage{lipsum}
\usepackage{nameref}
\usepackage{hyperref}
\usepackage{cleveref}

\usepackage[export]{adjustbox}
\usepackage{subscript}
\usepackage{titlesec,setspace}

\newcommand{\disp}[1]{Eq. (\ref{#1})}

\newcommand{\onlinecite}[1]{\hspace{-1 ex} \nocite{#1}\citenum{#1}}
\newcommand{\refdisp}[1]{Ref. [\onlinecite{#1}]}
\newcommand{\figdisp}[1]{Fig. \ref{#1}}

\newcommand{\beq}{\begin{eqnarray}}
\newcommand{\eeq}{\end{eqnarray}}

\title{Interaction-driven Spontaneous Ferromagnetic Insulating States with Odd Chern Numbers}

\author[1]{Peizhi Mai}
\author[1]{Edwin W. Huang}
\author[2,3]{Jiachen Yu}
\author[3,4,5]{Benjamin E. Feldman}
\author[1,*]{Philip W. Phillips}

\affil[1]{Department of Physics and Institute of Condensed Matter Theory, University of Illinois Urbana-Champaign, Urbana, IL 61801, USA}
\affil[2]{Department of Applied Physics, Stanford University, Stanford, CA 94305, USA}
\affil[3]{Geballe Laboratory of Advanced Materials, Stanford, CA 94305, USA}
\affil[4]{Department of Physics, Stanford University, Stanford, CA 94305, USA}
\affil[5]{Stanford Institute for Materials and Energy Sciences, SLAC National Accelerator Laboratory, Menlo Park, CA 94025, USA}
\affil[*]{dimer@illinois.edu}

\begin{abstract}
Motivated by recent experimental work on moir\'e systems in a strong magnetic field, we compute the compressibility as well as the spin correlations and Hofstadter spectrum of spinful electrons on a honeycomb lattice with Hubbard interactions using the determinantal quantum Monte Carlo method.  While the interactions in general preserve quantum and anomalous Hall states, emergent features arise corresponding to an antiferromagnetic insulator at half-filling and other incompressible states following 
the Chern sequence $\pm (2N+1)$.  These odd integer Chern states exhibit strong ferromagnetic correlations and arise spontaneously without any external mechanism for breaking the spin-rotation symmetry. Analogs of these magnetic states should be observable in general interacting quantum Hall systems.  In addition, the interacting Hofstadter spectrum is qualitatively similar to the experimental data at intermediate values of the on-site interaction. 
\end{abstract}

\begin{document}

\flushbottom
\maketitle

\section*{Introduction}

The hallmark of non-trivial topology of filled bands is a quantized Hall conductance, an integer multiple of the quantum of conductance.  The integer is set by the Chern number\cite{thouless}.   Typically, deviations from integer Chern numbers indicate that electron-electron interactions are important, as in the fractional quantum Hall effect.  However, there are several examples of physical systems in which interactions dominate but the Chern number is still an integer.  One such mechanism that involves spin polarization and its generalizations is quantum Hall ferromagnetism\cite{Sondhi,manfra1997,arovas1999,ezawa2002,odd1}. Moir\'e systems in a magnetic field provide a second example in which symmetry-broken quantum Hall insulators appear at high magnetic fields \cite{Spanton,Dean,kimbutterflied,geimbutterflied,butterflied}.  
The extent to which these phenomena are generic beyond graphene-based systems and independent of lattice geometry is unknown.

Motivated by these phenomena, we report here a series of insulating states on the honeycomb and square lattice which are driven by interactions. The series we report has odd integer Chern numbers, $\pm 1, 3, 5, \cdots$. An analysis of the spin correlations suggests that the spin rotation symmetry is spontaneously broken resulting in ferromagnetism. Our observations here add intrigue to the mixed role played by topology and interactions in 2D materials.  
Our simulations reveal that explicit single-particle symmetry breaking such as Zeeman splitting is not required and the full ferromagnetic sequence arises spontaneously in a general bipartite lattice.  

The evolution of electronic states in a perpendicular magnetic field has a long history.
Because a magnetic field preserves the crystal momentum, the single-particle energy spectrum for non-interacting electrons is easily obtained\cite{peierls,harper,hofstadter} by replacing the momentum ${\bf p}$ with ${\bf p}-e{\bf A}/c$ where ${\bf A}$ is the magnetic vector potential, $e$ the electron charge and $c$ the speed of light. In 2D, the resultant Hofstadter\cite{hofstadter} spectrum adequately describes the evolution of the tight-binding electronic states as a function of the magnetic flux.  Hidden in the wings of the underlying butterfly spectrum are gapped states indexed by Chern numbers which fix\cite{thouless} the quantization of the Hall conductance.  Moir\'e systems as in the case of MATBG offer a new route to engineering gaps in the electronic spectrum through the competition between band filling and the interaction energy\cite{Saito2,Park,Choi2,Ben}. As the twist angle controls\cite{bistritzer} the ratio of the kinetic to the potential energy and leads to a complete quenching of the kinetic energy at the magic angle, moir\'e systems in a magnetic field offer the ultimate playground for studying the physics from the interplay between strong correlation and magnetic field. With the kinetic energy quenched, moir\'e systems encode the evolution of the Hofstadter spectrum in the presence of strong interactions. This is currently an unsolved non-trivial problem.

This problem is complicated by the fact that the simple replacement of the momentum by ${\bf p}-e{\bf A}/c$ fails in the presence of interactions because interactions in general mix crystal momenta as in the case of the Hubbard interaction.  Consequently, while theoretical efforts have addressed certain limits of the interacting Hofstadter problem\cite{gudmundsson_effects_1995, pfannkuche_quantum_1997, doh_effects_1998, czajka_hofstadter_2006, apalkov_gap_2014, mishra_effects_2016, andrews_fractional_2020, tomD,Santos,Santos2,Oskar,Bernevig,DMFT1,DMFT2,HF1,HF2,HF3,HF4,ED1,ED2}, no analytical method exists to determine the complete spectrum in a magnetic field in the presence of interactions.  Nonetheless, this is an urgent problem in condensed matter physics given the plethora of experiments on MATBG and related systems that are focused on revealing the low-energy physics resulting from the interplay between a magnetic field and strong correlation.

Theoretically, one has three options: 1) phenomenology, 2) some type of mean-field theory, dynamical\cite{DMFT1,DMFT2} or otherwise\cite{HF1,HF2,HF3,HF4} or 3) serious numerics which so far have been limited to exact diagonalization\cite{ED1,ED2} on few-particle systems.  We pursue the last option in this paper as no benchmarks have been established for even the simplest model of interacting electrons on any of the lattices relevant to either MATBG or the transition metal dichalcogenide systems. 
We focus on spinful fermions primarily on a honeycomb lattice including only nearest neighbor hopping and Hubbard interactions under an external magnetic field, and perform a determinantal quantum Monte Carlo (DQMC) simulation for all densities and magnetic fluxes. DQMC is an unbiased and numerically exact method to capture the full quantum fluctuations for correlated systems.   In a prior work with Hubbard interactions, features such as the local compressibility and other thermodynamic quantities were calculated using DQMC as a function of the magnetic flux for the square lattice\cite{tomD} and no ferromagnetism was reported.  We focus here on a honeycomb lattice as it is closer to the underlying geometry of most existing moir\'e systems.  In general, we find that the interactions preserve the integer quantum and anomalous Hall states of the non-interacting system.  However, the interactions do generate an antiferromagnetic insulating state at half-filling, as expected, and also emergent interaction-driven insulating states in both of the honeycomb and square lattices.  The Chern sequence for these states is $\pm (2N+1)$. All such states exhibit strong ferromagnetic correlations.  This represents a numerically exact evidence for such interaction-driven states in full density region based on the Hubbard interaction.

\section*{Results}
\subsection*{Non-interacting quantum Hall effects}
\begin{figure*}[ht]
    \centering
    \includegraphics[width=1.0\textwidth]{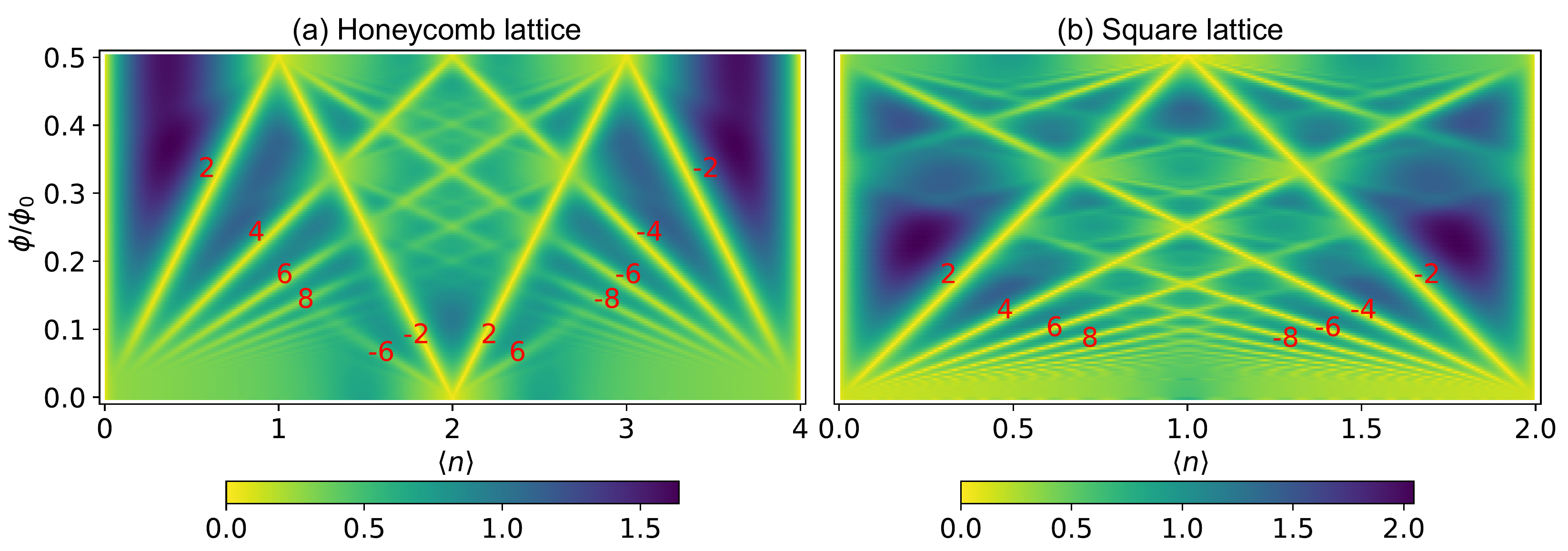}
    \caption{{\bf Non-interacting compressibility for square and honeycomb lattice.} Non-interacting compressibility as a function of magnetic flux $\phi$ and electron density $\langle n \rangle$ for (a) honeycomb lattice ($N_{\text{site}} = 36\times36\times2$) and (b) square lattice ($N_{\text{site}} = 40\times40$) at $\beta/t=20$. The leading quantum Hall states are labeled with the corresponding Chern number. }
    \label{fig:hcvssq}
\end{figure*}

To begin with, we present the non-interacting charge compressibility $\chi=\partial \langle n \rangle/\partial \mu$ in \figdisp{fig:hcvssq} as a function of magnetic flux and electron density and compare with the results for the square lattice at $\beta=20/t$. In both the square and honeycomb lattices,  particle-hole symmetry obtains as both are bipartite and the model contains only nearest neighbor hopping. The straight lines in \figdisp{fig:hcvssq} correspond to solutions to the Diophantine equation \cite{thouless}
\beq
\langle n\rangle=r \frac{\phi}{\phi_0}+s,
\eeq
in which $\langle n\rangle=\langle N_\text{e}\rangle/N_c$ ($N_c$ is the number of unit cells), $r$ is an integer given by the inverse slope of the straight lines and $s$ is the offset given by the intercept. $r$ defines the Chern number. We have chosen to plot the filling from $[0,4]$ to take into account the spin and sublattice degeneracy in the honeycomb lattice but from $[0,2]$ for the square lattice in which only a spin degeneracy exists. Hence, there is only a factor of $2$ in translating the densities between the two systems.  In both  \figdisp{fig:hcvssq} panels (a) and (b), the Diophantine lines\cite{thouless} starting from the bottom left (right) corners have $r=\pm 2N$ (the factor of 2 accounts for spin degeneracy) with $N=1,2\cdots$ corresponding to spin-unpolarized quantum Hall states, while in only \figdisp{fig:hcvssq}(a), the lines that start from half-filling ($\langle n\rangle=2$) at zero-field and have $r=\pm4(N+1/2)$ 
(the factor of 4 accounts for spin and sub-lattice degeneracy) with $N=0,1,2\cdots$ indicating the anomalous quantum Hall effect\cite{semenoff,Firsov,Kim}.  Thus, the honeycomb lattice is more closely aligned to the physics observed in MATBG than is the square lattice.

\subsection*{Turning on interactions}
\begin{figure*}[ht]
    \includegraphics[width=1.0\textwidth]{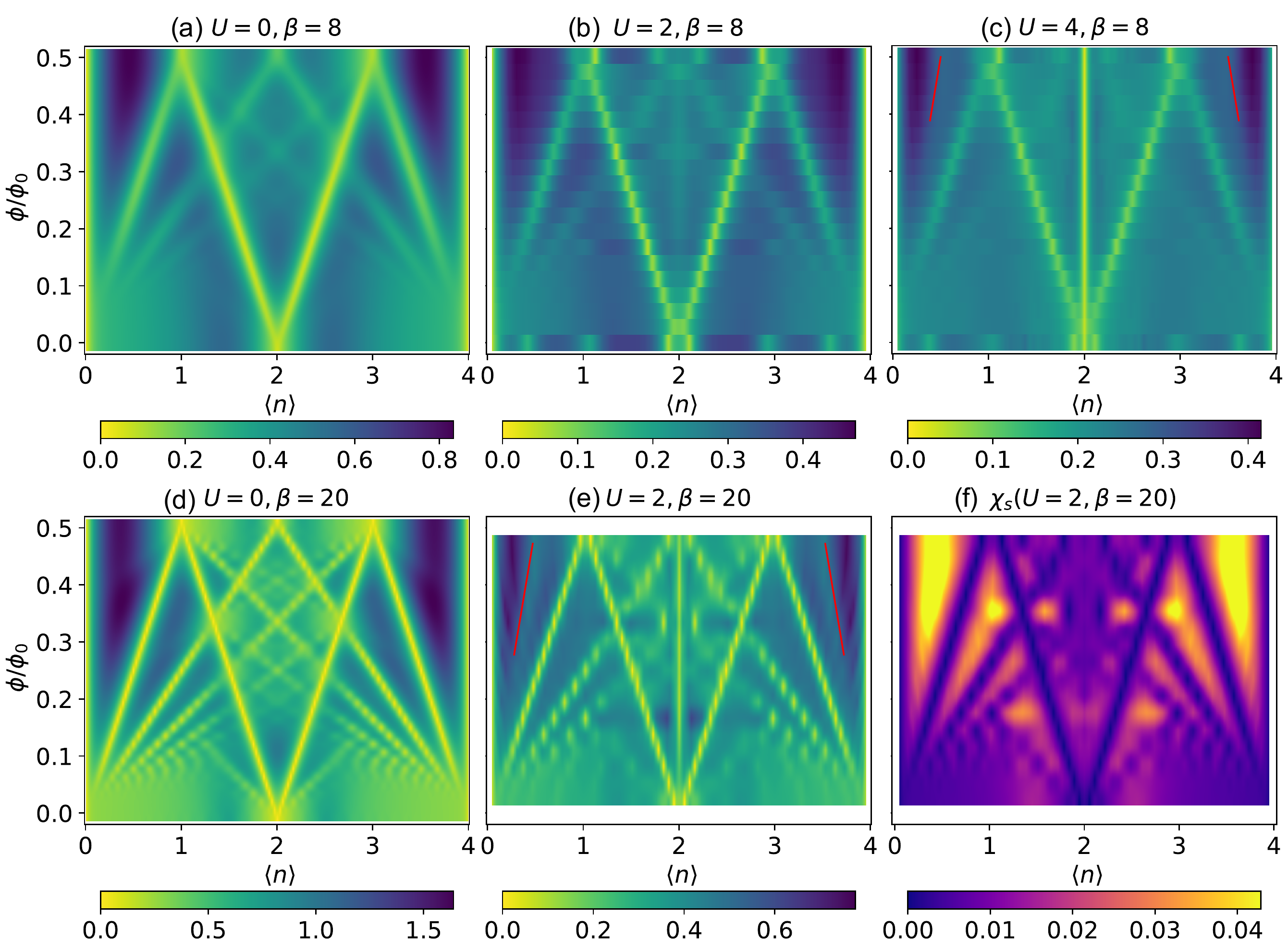}
    \caption{{\bf Compressibility and spin correlation in the presence of interactions.} Compressibility as a function of magnetic flux and electron density in an $N_{\text{site}}=36\times36\times2$ cluster for (a,d) $U/t= 0$; an $N_{\text{site}}=6\times6\times2$ cluster for (b,e) $U/t=2$ and (c) $U/t=4$. The first row is for the inverse temperature $\beta=8/t$ and the second row for $\beta=20/t$.  Panel (f) shows the corresponding spin correlation at $U/t=2$ and $\beta=20/t$. It has a reverse color bar compared to those of the charge compressibilty.}
    \label{fig:varyU}
\end{figure*}

Next, we explore how interactions change this pattern in the honeycomb lattice. We use DQMC to calculate the compressibility,
\beq
\chi=\beta\chi_c=\frac{\beta}{N}\sum_{{\bf i},{\bf j}}\left[\langle n_{\bf i} n_{\bf j}\rangle - \langle n_{\bf i}\rangle \langle n_{\bf j}\rangle \right],
\eeq
in the presence of Hubbard interactions, where $\chi_c$ is the charge correlation function. Due to the Fermionic sign problem, we are only able to calculate the compressibility for the full density and flux region for a system size $N_{\text{site}}=6\times6\times2$ with an interaction strength up to $U/t=4$ and temperature as low as $T/t=0.125$ (or $\beta=8t^{-1}$).   In the first row of \figdisp{fig:varyU}, as $U$ increases, the non-interacting lines in the compressibility are softened and a middle vertical line appears as a single-particle gap develops. At the largest $U/t=4$, we can still observe the dominant and sub-dominant lines, indicating the resilience of the non-interacting quantum Hall effect against interactions. Note that the lines not merging at $\phi/\phi_0=0,0.5$ in \figdisp{fig:varyU}(b,c) is due to strong finite size effects (see Supplementary Figure 1 and 2) and thus disregarded in \figdisp{fig:varyU}(e,f). Also of note is the emergence of a new feature at $\langle n\rangle=2$ for $U/t=4$.  This corresponds to a dip in the density of states, a precursor to the Mott gap\cite{Assaad2,Ostmeyer}.  The second row of these figures corresponds to a simulation at the lower temperature of $\beta=20/t$ for $U=0$ and $U/t=2$.  \figdisp{fig:varyU}(d) shows a sharpening of the Diophantine features as the temperature is lowered by more than a factor of two to $\beta=20/t$. At $U/t=2$, the vertical line at $\langle n\rangle=2$ possibly indicates a gap opening for finite field. At this temperature, even for such a modest value of $U$, the suppression of the density of states is evident and the corresponding insulator is antiferromagnetic (see Supplementary Figure 6). Also of note is the state indicated by the solid red line in \figdisp{fig:varyU}(c,e). This line evolves as a function of the magnetic flux with a slope of unity. Such a state is absent from the non-interacting sequence as it has a Chern number of $\pm 1$. Notably this state is visible in the map of the spin correlation (\figdisp{fig:varyU}(f)), 
\beq
\chi_s=\sum_{\bf r}S({\bf r})=\frac{1}{N}\sum_{{\bf i},{\bf r}}\langle S^z_{\bf i} S^z_{{\bf i}+{\bf r}}\rangle.\label{spin}
\eeq
 In fact, the light features in \figdisp{fig:varyU}(f) reveal that these possibly incompressible states all have a markedly enhanced spin correlation with distinct slopes as a function of magnetic flux and density which differ from those of the non-interacting system even at much lower temperatures (see Supplementary Figure 4). It is the origin of these states that is the principal focus of this paper.

\begin{figure*}[ht]
    \centering
    \includegraphics[width=1.0\textwidth]{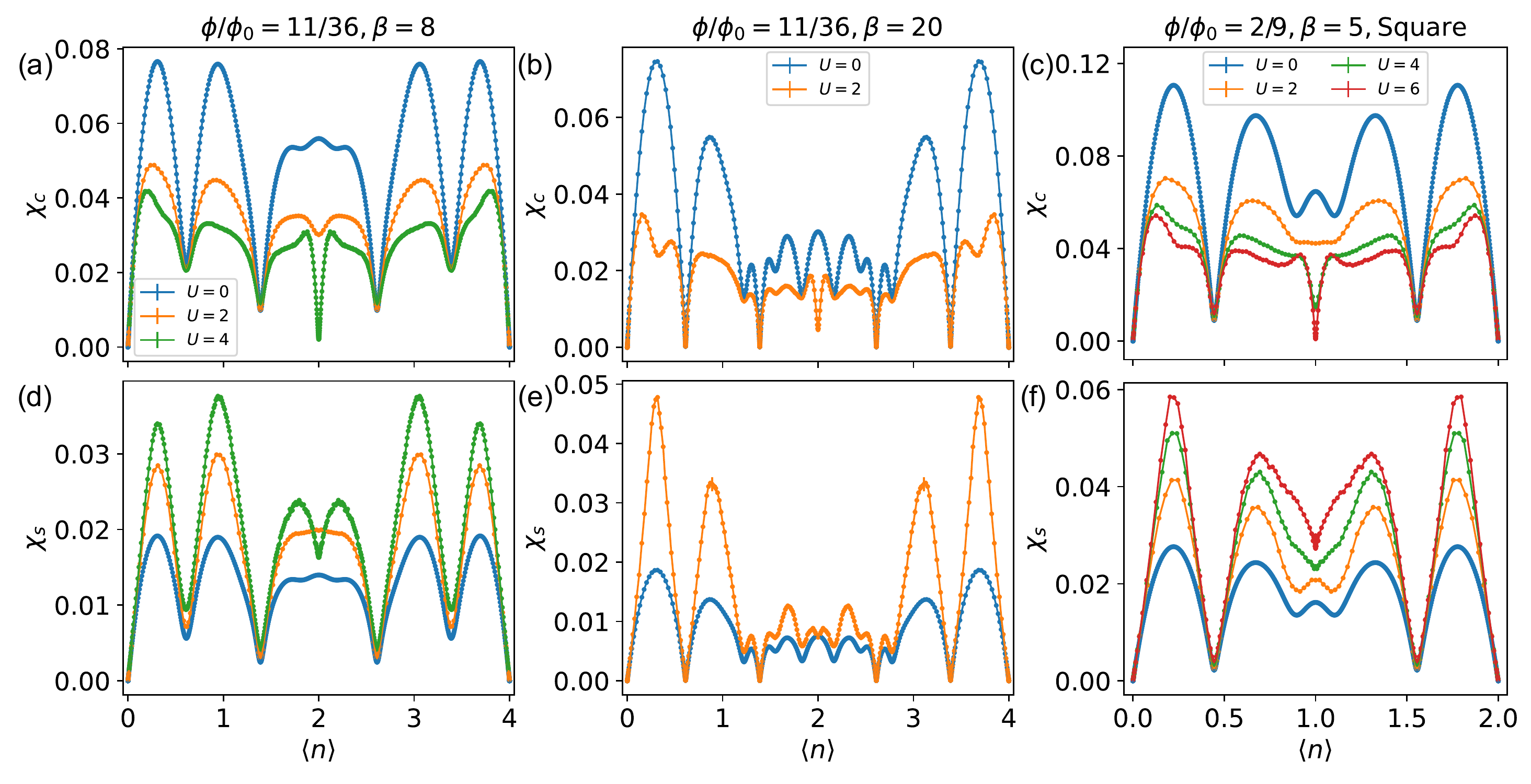}
    \caption{{\bf Compressibility and spin correlation at a specific magnetic flux.} The first row shows the charge correlation at varying interaction strength with different fixed parameter sets. Panel (a) and (b) are for the honeycomb lattice at $\phi/\phi_0=11/36$, while panel (c) is for the square lattice at $\phi/\phi_0=2/9$. The corresponding temperatures are (a) $\beta=8/t$, (b) $\beta=20/t$,  and (c) $\beta=5/t_\text{sq}$ ($t_{sq}$ is the nearest neighbor hopping for the square lattice). The second row shows the corresponding spin correlation and shares the same legend as does the first row.}
    \label{fig:flux11}
\end{figure*}

We gain further insight into the possible emergence of incompressible states by taking slices through the charge correlation, $\chi_c=\chi/\beta$, at particular values of the magnetic flux.  In \figdisp{fig:flux11}(a-c), we show the charge correlation explicitly at $\phi/\phi_0=11/36$, chosen to avoid finite-size effect (see Supplementary Figure 1 and 2) for the honeycomb lattice and $\phi/\phi_0=2/9$ for the square lattice.  For the honeycomb lattice,  \figdisp{fig:flux11}(a,b) illustrates that as the temperature is lowered from $\beta=8/t$ to $\beta=20/t$, the dual-dip feature for $U=0$ in the vicinity of $\langle n\rangle=2$ gives rise to a full quantum Hall state. With the interaction increasing to $U=4t$, this dual-dip feature contains a depression precisely at $\langle n\rangle=2$.  This is the incompressible Mott gap at $\langle n\rangle=2$ \cite{Assaad2,Ostmeyer}. Panel \figdisp{fig:flux11}(c) displays the analogous trend for the square lattice. Away from half-filling, we find several emergent states which have a suppressed charge correlation indicating the possible onset of a gap. The states occur around fillings of $\langle n\rangle =11/36$ and $11/12$. The same is true of their particle-hole equivalents. It is precisely the first of such states that is highlighted in red in  \figdisp{fig:varyU}(e). The analogous states are also present for the square lattice in \figdisp{fig:flux11}(c).   All such dips in $\chi_c$ are enhanced as the interaction strength increases. Further, such behavior persists even as the system size increases (see Supplementary Figure 3). To uncover the possible cause of these states, we focus on the spin susceptibility $\chi_s$. For $U=0$, $\chi_c=4\chi_s$. However, the second row of \figdisp{fig:flux11} shows that whenever the charge correlation exhibits an interaction-driven dip, the spin correlation shows a peaked structure. \figdisp{fig:flux11}(e) shows that as the temperature is lowered, the peak of the spin correlation increases at the fillings where the charge correlation develops a dip.

\begin{figure*}[ht]
    \centering
    \includegraphics[width=1.0\textwidth]{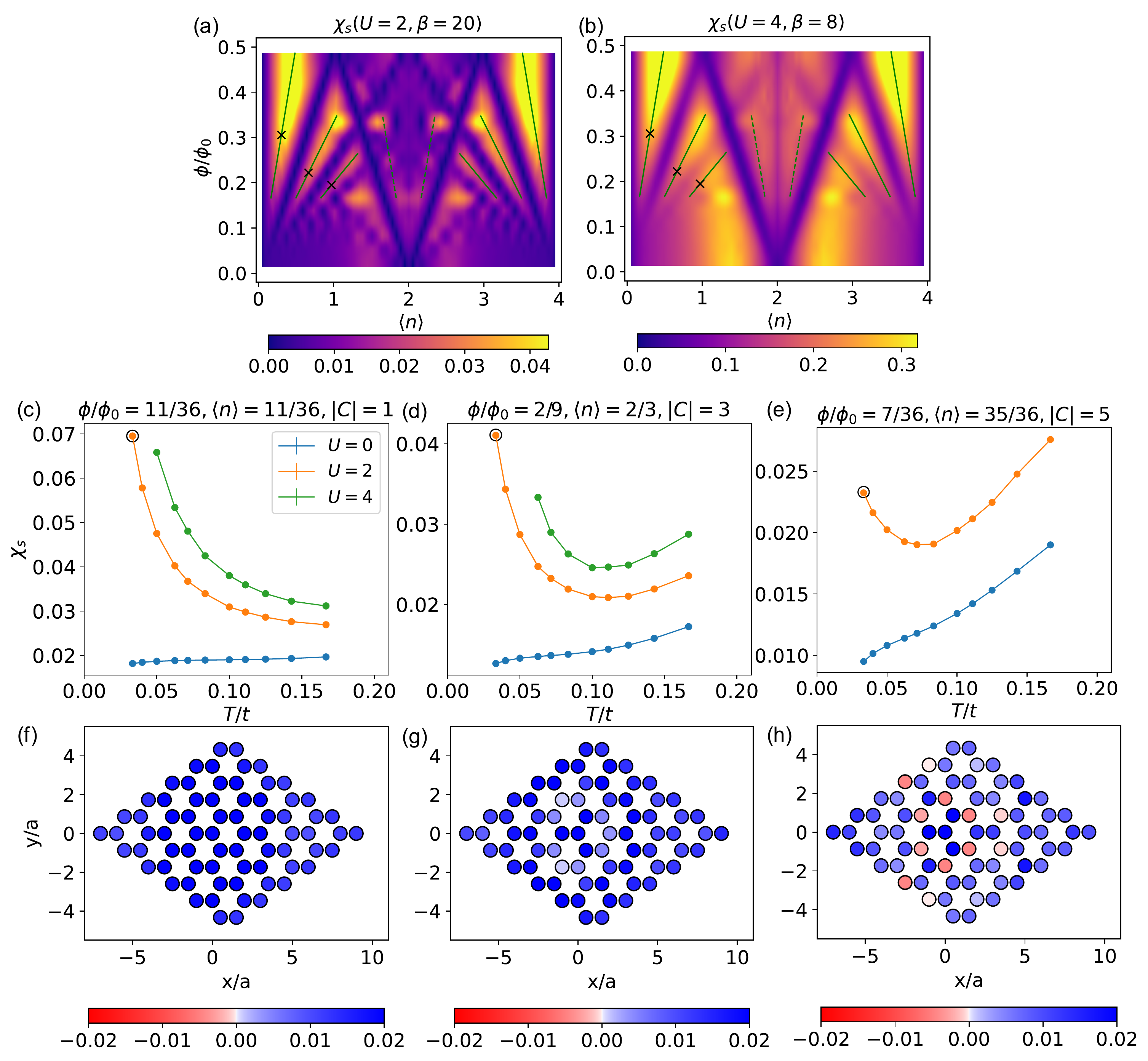}
    \caption{{\bf Spin correlation and ferromagnetic states.} The first row shows the spin correlation at (a) $U/t=2, \beta=20/t$ and (b) $U/t=4, \beta=8/t$, with the green lines depicting the odd integer Chern states aligned with the ridge of the spin correlation. The second row shows the spin correlation for selected points (marked at each Chern state in panels (a) and (b)) as a function of temperature under different interaction strengths. The third row presents the real-space zero-frequency spin susceptibility pattern for the circled points in the corresponding second row (at $U/t=2$ and the lowest temperature $\beta=30/t$).
    }
    \label{fig:FM}
\end{figure*}

Now we look at the full density- and magnetic-flux-dependent spin correlation $\chi_s$ for $U/t=2$ and $U/t=4$ in \figdisp{fig:FM}(a) and (b) respectively at their lowest temperatures. Straight lines with inverse slope corresponding to Chern number $C=\pm1, 3, 5$ are plotted and found to be aligned with the ridges of the spin correlation. We choose one representative point (not the brightest) at each line and study its temperature evolution at different $U$, presented in \figdisp{fig:FM} (c-e). In all cases, when $U=0$, the spin correlation deceases along with temperature. However, for a finite $U$, the spin correlation blows up as temperature decreases (below a critical temperature for \figdisp{fig:FM} (d) and (e)). The ultimate
spin state is revealed from a spatial map of the real-space static spin susceptibility:
\beq
S({\bf r},\omega=0)=\frac{1}{N}\int_0^{\beta}\sum_{{\bf i}}\langle S^z_{{\bf i}+{\bf r}}(\tau) S^z_{\bf i}(0) \rangle d\tau,\label{spinsus}
\eeq
presented in \figdisp{fig:FM} (f-h), at $U/t=2$ and the lowest temperature ($\beta=30/t$ as circled in \figdisp{fig:FM} (c-e) respectively). This quantity is more sensitive in detecting fluctuating order at finite temperature than the zero-time spin correlation\cite{Mai}. The color map signifies positive spin correlation across the lattice relative to the site at the origin. Such same-sign correlations are indicative of ferromagnetism. \figdisp{fig:FM}(f) for a Chen number $C=1$ exhibits strong ferromagnetic susceptibility. \figdisp{fig:FM}(g) with a Chen number $C=3$ also displays a clear ferromagnetic pattern. \figdisp{fig:FM}(h) corresponding to Chen number $C=5$ reveals an evident tendency towards ferromagnetism at lower $T$, though not fully ferromagnetic as in the other cases. In addition, we also expect the $C=\pm 1$ ferromagnetic states to exist in the middle of \figdisp{fig:FM} (a,b) (as depicted by the dashed line) with extrapolation to $\langle n\rangle=2$ at zero flux. But limited by the sign problem, we have yet been not able to investigate low enough temperature to unearth a clear ferromagnetic pattern at these densities.

The full picture is now apparent. The charge dips and enhanced spin correlations, which have no counterpart in the non-interacting system in \figdisp{fig:varyU}, correspond to ferromagnetic insulators with odd integer Chern numbers.   Both the spin correlations and magnitude of the charge gap are enhanced as the temperature is lowered.  The same trend holds for the square lattice as is evident from \figdisp{fig:flux11}(c,f). This behavior matches expectations from quantum Hall ferromagnetism, but here we show that local interactions are sufficient to induce odd Chern integer states on both the honeycomb and square lattices.  We are led to the conclusion that such insulating states are generically present in bipartite lattices with interactions with no need for fine-tuning or single-particle splitting.   
While there is some indication of charge ordering (see Supplementary Figure 5), it is not compelling at this level of study and hence we leave this for a future publication. 

\begin{figure}
    \centering
    \includegraphics[width=0.5\textwidth]{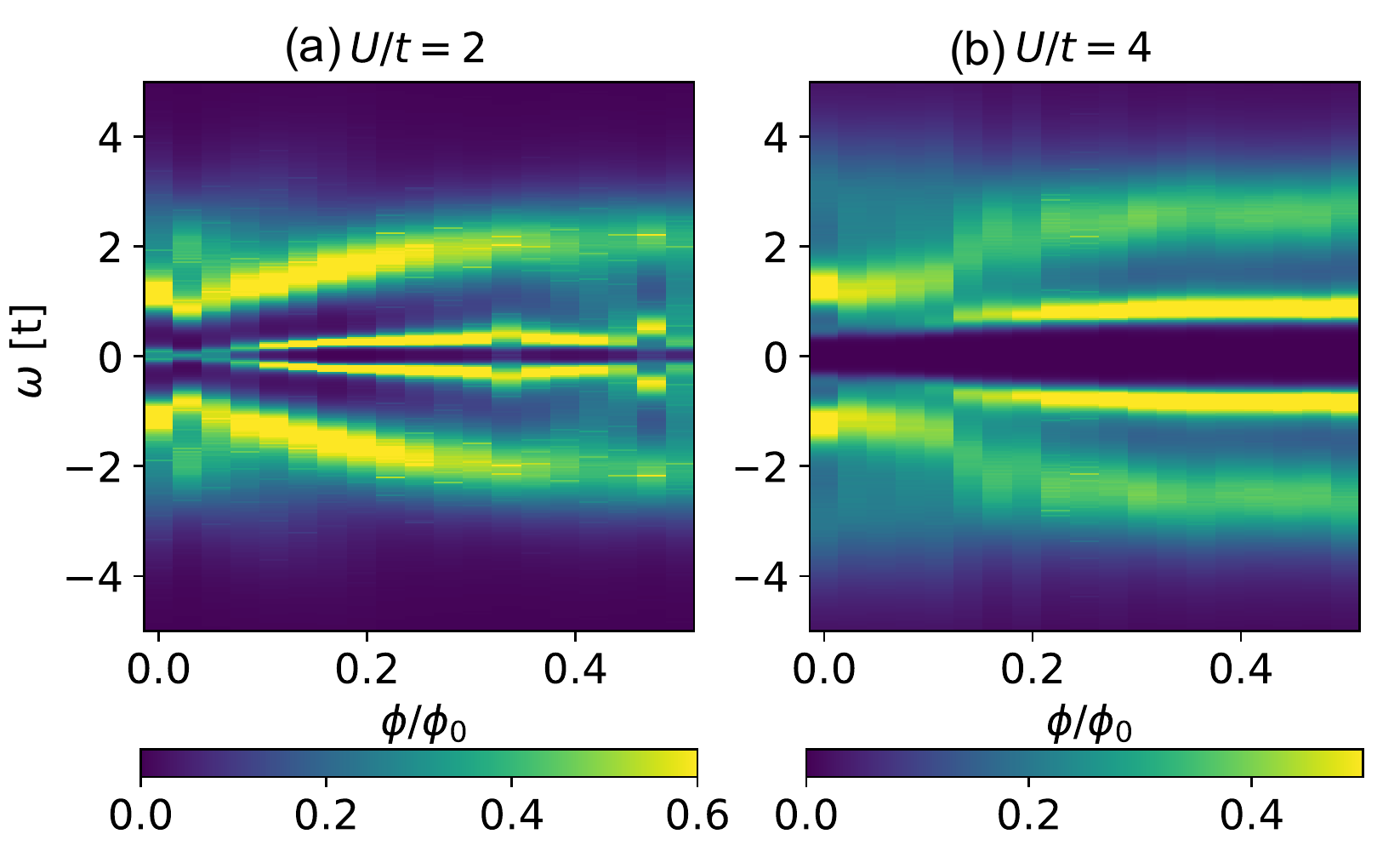}
    \caption{{\bf Density of states at half-filling.} The density of states at half-filling ($\langle n\rangle=2$) and $\beta=30/t$ for (a) $U/t=2$ and (b) $U/t=4$.  }
    \label{fig:DOS}
\end{figure}

We finally display the benchmark calculation of the Hofstadter spectrum as defined by the local density of states, the quantity directly measured experimentally.  Our focus in \figdisp{fig:DOS} is at half-filling and $U/t=2,4$, obtained from constructing an analytic continuation with DEAC on the DQMC local Green function. The comparison between DEAC and the analytical result at $U=0$ (see Supplementary Figure 7) gives us some idea about the resolution of DEAC and offers a guide as to how to interpret the interacting system results. Since there is no sign problem at half-filling, we are able to conduct the calculation at a low temperature ($\beta=30/t$).  Panels \figdisp{fig:DOS}(a,b) show how the antiferromagnetic gap comes into full view by $U/t=4$ and some hint of it appears already at the modest value of $U/t=2$ only with finite magnetic field. The gap at $U/t=2$ is most likely to be of the Slater type\cite{Gull_2008,Schafer} because the interaction strength is only around $1/3$ of the bare bandwidth and the insulating state appears at much lower temperature than that required for the formation of antiferromagnetic correlation. On the other hand, the gap at $U/t=4$ is closer to a Mott gap because it is established at a much high temperature (shown in \figdisp{fig:varyU}(c)), consistent with previous studies on the Hubbard model in honeycomb lattice\cite{Assaad2,Ostmeyer}.  While the corresponding experimental figure is at variable filling\cite{Ben}, which is inaccessible because of the sign problem, the overall features qualitatively reproduce the experimental results for moderate values of $U/t=2$.

\section*{Discussion}

We have studied here the evolution of the excitation spectrum of a Hubbard-interacting electron gas in the presence of a strong perpendicular magnetic field  on bipartite lattices.  We have shown that while the interactions preserve the non-interacting integer quantum and anomalous Hall states, new states do emerge from the interactions.  In addition to the antiferromagnetic gap at half-filling, we have discovered a series of odd-integer Chern insulating ferromagnetic states which exhibit enhanced positive spin correlations as the temperature is lowered.  In light of ferromagnetism as the underlying cause of the insulating states, that the Chern number is odd is easily understood.   Our work suggests ferromagnetism is generic requiring just modest magnetic fields and strong interactions to generate the full sequence of odd-integer states.  
Note while the sequence we observe departs from the middle or the edge of the band, an odd-integer sequence emanating from the $\langle n\rangle=1$ or $\langle n\rangle=3$ fillings on the honeycomb lattice would require spin-orbit coupling in the Hamiltonian, thereby generalizing the utility of this work. 

\section*{Methods}

We study the Hofstadter-Hubbard model on a honeycomb lattice,
\beq\label{Eq:HubbardBfield}
\begin{aligned}
    H&= -t\sum_{\langle{\bf i}{\bf j}\rangle\sigma} \exp(i \phi_{{\bf i},{\bf j}} ) c^\dagger_{{\bf i}\sigma}c^{\phantom\dagger}_{{\bf j}\sigma}   
    -\mu\sum_{{\bf i},\sigma} n_{{\bf i}\sigma} \\
    &+ U\sum_{{\bf i}}(n_{{\bf i}\uparrow}-\frac{1}{2})(n_{{\bf i}\downarrow}-\frac{1}{2}),
\end{aligned}
\eeq
where $t$ represents the nearest neighbor hopping; $c^\dagger_{{\bf i}\sigma}$ ($c_{{\bf i}\sigma}$) creates (annihilates) an electron with spin $\sigma$ at site ${\bf i}$, $\mu$ is the chemical potential, $U$ is the on-site interaction. Due to the presence of a uniform magnetic field, we use the Peierls substitution\cite{peierls} to introduce the phase through the flux threading,
\beq
\phi_{{\bf i},{\bf j}}=\frac{2\pi }{\phi_0} \int_{r_{\bf i}}^{r_{\bf j}} {\bf A}\cdot d{\bf l},
\eeq
where $\phi_0=h/e$ in the hopping term is a result of the quantized magnetic field and the integration is over the straight line path from site ${\bf i}$ to ${\bf j}$. 

We simulate this Hamiltonian \disp{Eq:HubbardBfield} on a finite cluster $N_{\text{site}}=2L^2$. The honeycomb lattice contains two sub-lattices, which explains the factor of $2$, with lattice constant $a=1$ and $L$ the number of site along either lattice basis respectively for each sub-lattice. We adjust the modified periodic boundary conditions in \refdisp{Assaad} to the honeycomb lattice. To obtain a single-value wave function requires the flux quantization condition $\phi/\phi_0=n_f/N_c$ with $n_f$ an integer and $N_c=L^2$ the number of unit cells. The symmetric gauge ${\bf A}=(x\hat{y}-y\hat{x})B/2$ is chosen for this calculation.

We apply DQMC\cite{Blankenbecler,Hirsch,White} to this model \disp{Eq:HubbardBfield} and calculate the compressibility and Green function. The jackknife resampling is used to estimate the standard error of the mean as error bars in DQMC results. With the local Green function, we compute the local density of states using the recently developed Differential Evolution for Analytic Continuation (DEAC) \cite{DEAC}. 

\section*{Data Availability} The data for this study is available at \url{https://zenodo.org/record/7608167#.Y-AxyezML6g}.

\section*{Code Availability} The DQMC code used for this project can be obtained at \url{https://github.com/edwnh}.

\section*{Acknowledgements} 

BEF, PM, PWP, and JY acknowledge  support for the computation and conception of this project from Quantum Sensing and Quantum Materials (QSQM), an Energy Frontier Research Center funded by the US Department of Energy (DOE), Office of Science, Basic Energy Sciences (BES), under award no. DE-SC0021238. E.W.H. was supported by the Gordon and Betty Moore Foundation EPiQS Initiative through the grants GBMF 4305 and GBMF 8691.  This work used an analysis of insulating states in MATBG funded through DMR-2111379 for his work on MATBG. PM thanks Nathan Nichols for the help on DEAC simulation. The DQMC calculation of this work used the Advanced Cyberinfrastructure Coordination Ecosystem: Services \& Support (ACCESS) Expanse supercomputer through the research allocation TG-PHY220042, which is supported by National Science Foundation grant number ACI-1548562\cite{xsede}.

\section*{Competing interests}
The authors declare no competing financial or non-financial interests.

\section*{Author Contributions}
E.~W.~H. and P.~M. developed the DQMC code. P.~M. carried out the calculations. All authors analyzed the results and wrote the manuscript. B.~E.~F. and P.~W.~P. supervised the project.

\bibliography{reference}

\newpage
\section*{Figure Legend}

\end{document}


\title{Interaction-driven Spontaneous Ferromagnetic Insulating States with Odd Chern Numbers: supplementary information}

\author{ Peizhi Mai$^{1}$,  Edwin W. Huang$^{1}$, Jiachen Yu$^{2,3}$, Benjamin E. Feldman$^{3,4,5}$, Philip W. Phillips$^{1,*}$ }

\affiliation{$^1$Department of Physics and Institute of Condensed Matter Theory, University of Illinois Urbana-Champaign, Urbana, IL 61801, USA}
\affiliation{$^2$Department of Applied Physics, Stanford University, Stanford, CA 94305, USA}
\affiliation{$^3$Geballe Laboratory of Advanced Materials, Stanford, CA 94305, USA}
\affiliation{$^4$Department of Physics, Stanford University, Stanford, CA 94305, USA}
\affiliation{$^5$Stanford Institute for Materials and Energy Sciences, SLAC National Accelerator Laboratory, Menlo Park, CA 94025, USA}
\email{dimer@illinois.edu}

\maketitle

\section*{Supplementary Note 1: Finite-size effect}
\begin{figure*}[ht]
    \centering
    \includegraphics[width=\textwidth]{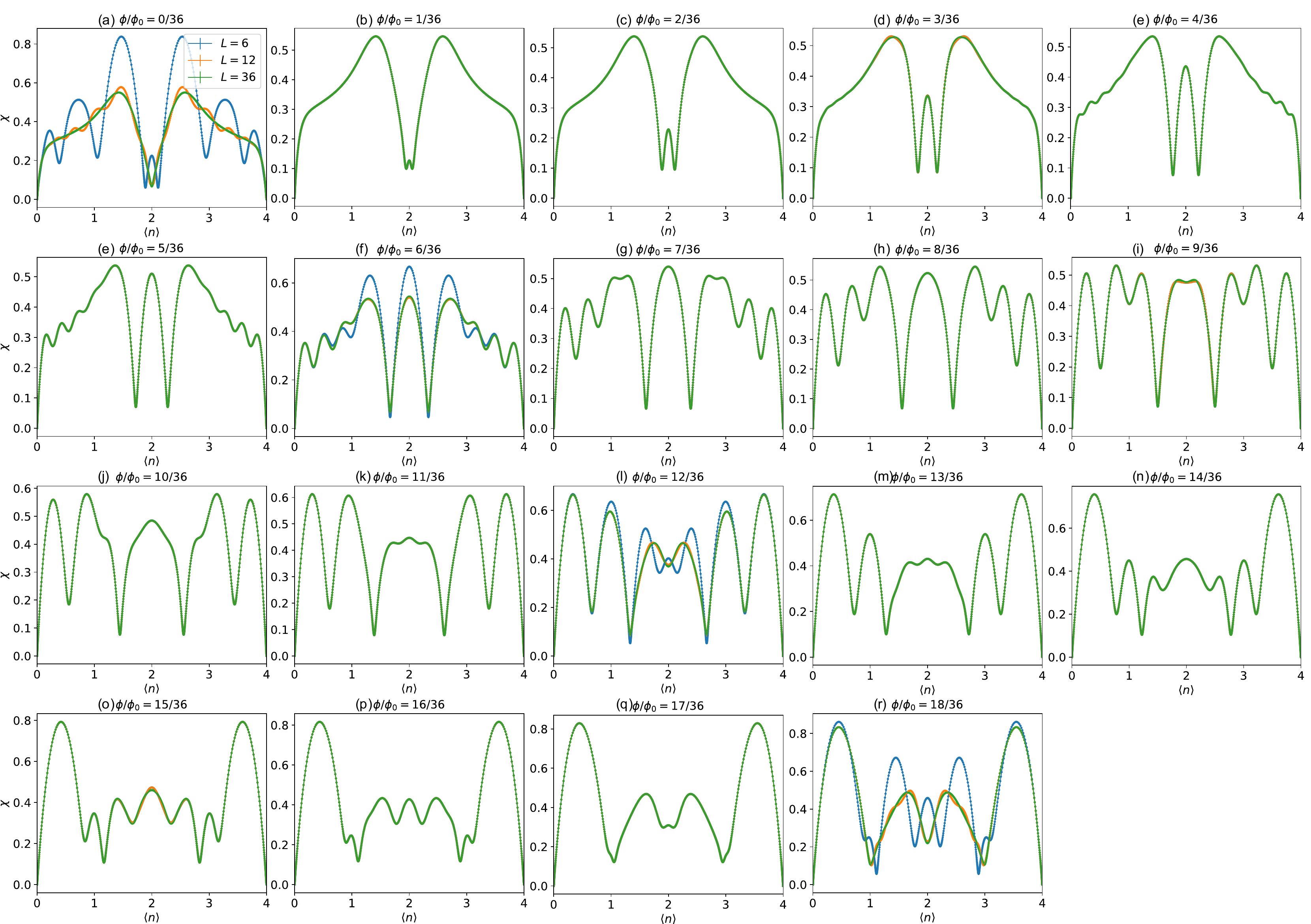}
    \caption{Non-interacting compressibility for several system sizes ($L=6, 12, 36$) under all different magnetic fluxes. The inverse temperature is $\beta=8/t$. All panels share the same legend. }
    \label{fig:b8FSE}
\end{figure*}

\begin{figure*}[ht]
    \centering
    \includegraphics[width=\textwidth]{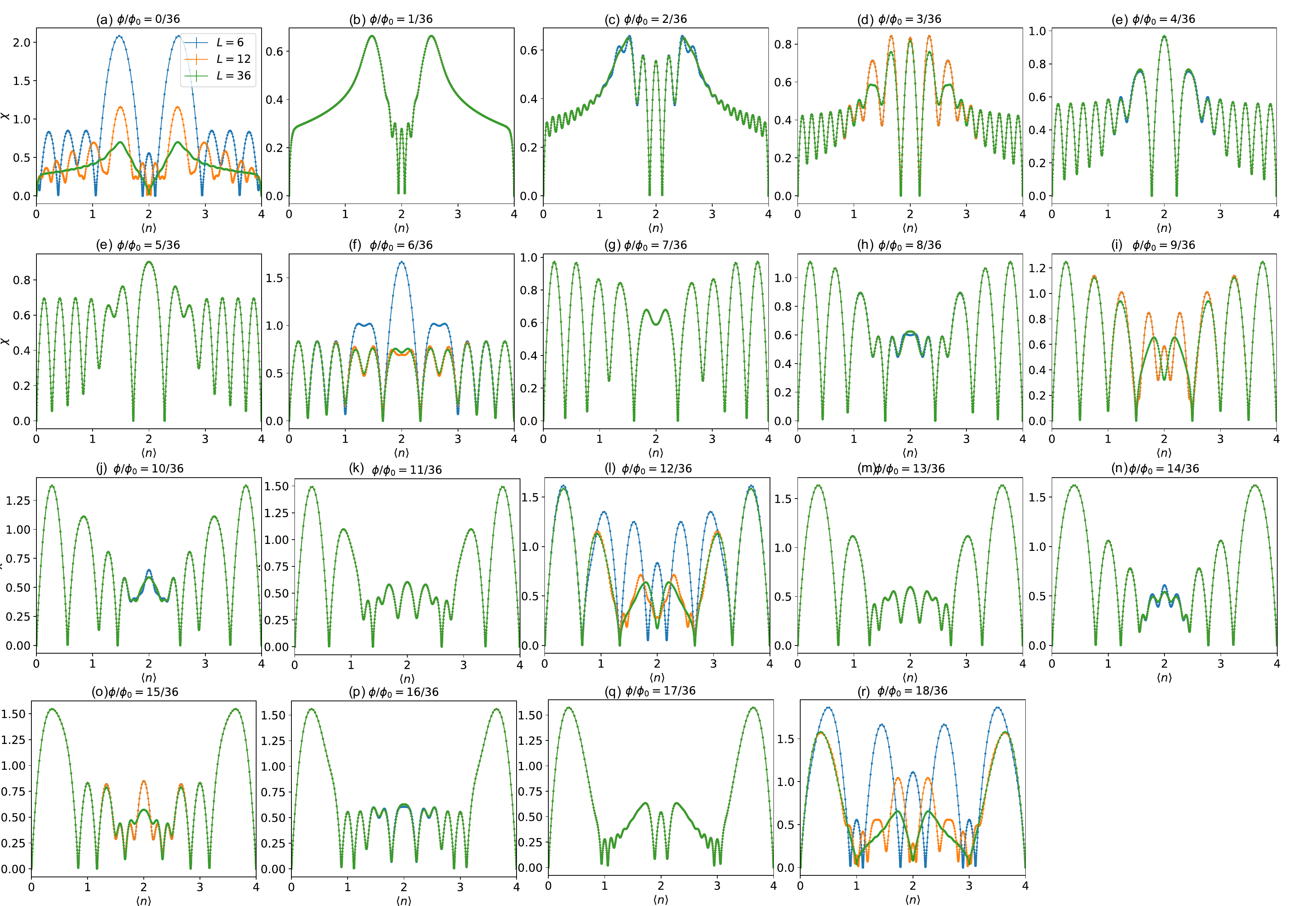}
    \caption{Non-interacting compressibility for several system sizes ($L=6, 12, 36$) under all different magnetic fluxes. The inverse temperature is $\beta=20/t$. All panels share the same legend. }
    \label{fig:b20FSE}
\end{figure*}

\begin{figure*}[ht]
    \centering
    \includegraphics[width=0.4\textwidth]{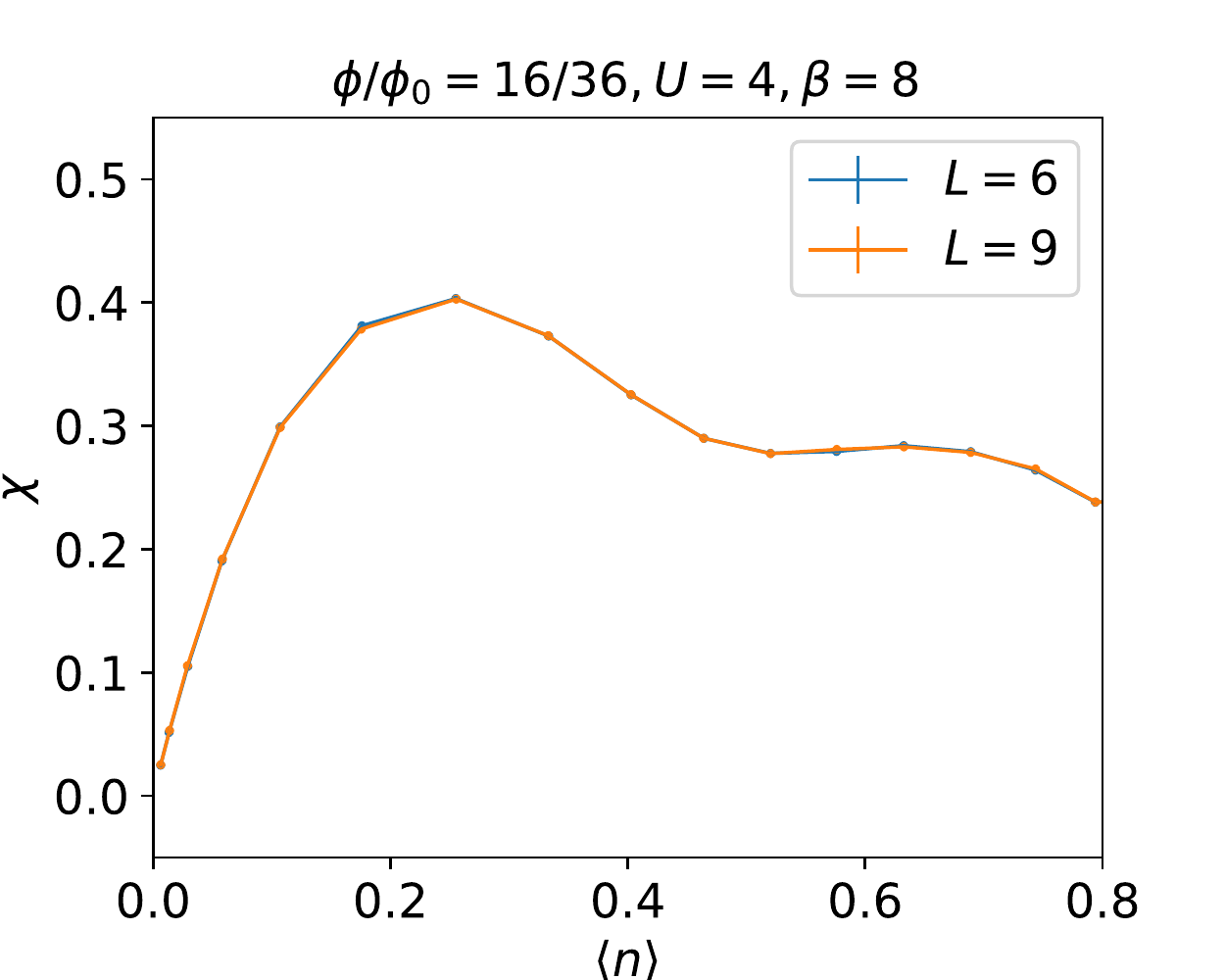}
    \caption{The incompressible region in the compressibility caused by the interaction for different cluster sizes $N_{\text{site}}=6\times6\times2$ and $N_{\text{site}}=9\times9\times2$. The fixed parameters are $\beta=8/t$, $U=4t$, $\phi/\phi_0=16/36$. Both curves collapse, indicating the incompressible state is not related to finite-size effects.}
    \label{fig:varys}
\end{figure*}

In the presence of interactions, we need to use determinantal quantum Monte Carlo (DQMC) to solve the Hamiltonian. The DQMC method suffers from the Fermionic sign problem which becomes worse for strong interaction, low temperature and large cluster sizes. We are restricted to a cluster size $N_{\text{site}}=6\times6\times2$. Therefore, we need to gauge finite-size effects to ensure that our conclusion from the $6\times6\times2$ cluster applies to the thermodynamic limit.

We show the finite-size effect in the non-interacting system in \figdisp{fig:b8FSE} ($\beta=8/t$) and \figdisp{fig:b20FSE} ($\beta=20/t$) for all magnetic fluxes. At $\beta=8/t$, the compressibility endures strong finite-size effects for $\phi/\phi_0=0, 0.5$ as evidenced by the unphysical valleys which become flattened as the system size increases. This explains the "additional" state shown in $\phi/\phi_0=0, 0.5$ of Fig.~2(b,c) in the main text. This state is incompatible with the remaining part of the panel ($0<\phi/\phi_0<0.5$). In addition, for $\phi/\phi_0=n/6~(n=1,2)$, the finite-size effect is considerable, predominantly around $\langle n\rangle=2$. For other values of magnetic flux, there is barely any finite-size effect because the lines collapse among different system sizes. At $\beta=20/t$ in \figdisp{fig:b20FSE}, the finite-size effect gets stronger, as expected. Apart from $\phi/\phi_0=0, 0.5$, the considerable deviation from larger clusters now appears at $\phi/\phi_0=n/12~(n=1,2,3,4,5)$. Thus, in the main text, we fix $\phi/\phi_0=11/36$ and show the density-dependent compressibility and spin correlation to avoid the finite-size effect regime. Our conclusion with this value of the magnetic flux should apply to at least $36\times36\times2$ cluster size and therefore the thermodynamic limit.

To be rigorous, in \figdisp{fig:varys}, we show the incompressible state of the interacting system in the dilute region at different lattice sizes. The results overlay as expected for $\phi/\phi_0=16/36$ which illustrates that this emergent incompressible state with Chern number $C=-1$ is robust against increasing system size.

\clearpage
\section*{Supplementary Note 2: Non-interacting compressibility}

\begin{figure*}[ht]
    \centering
    \includegraphics[width=0.6\textwidth]{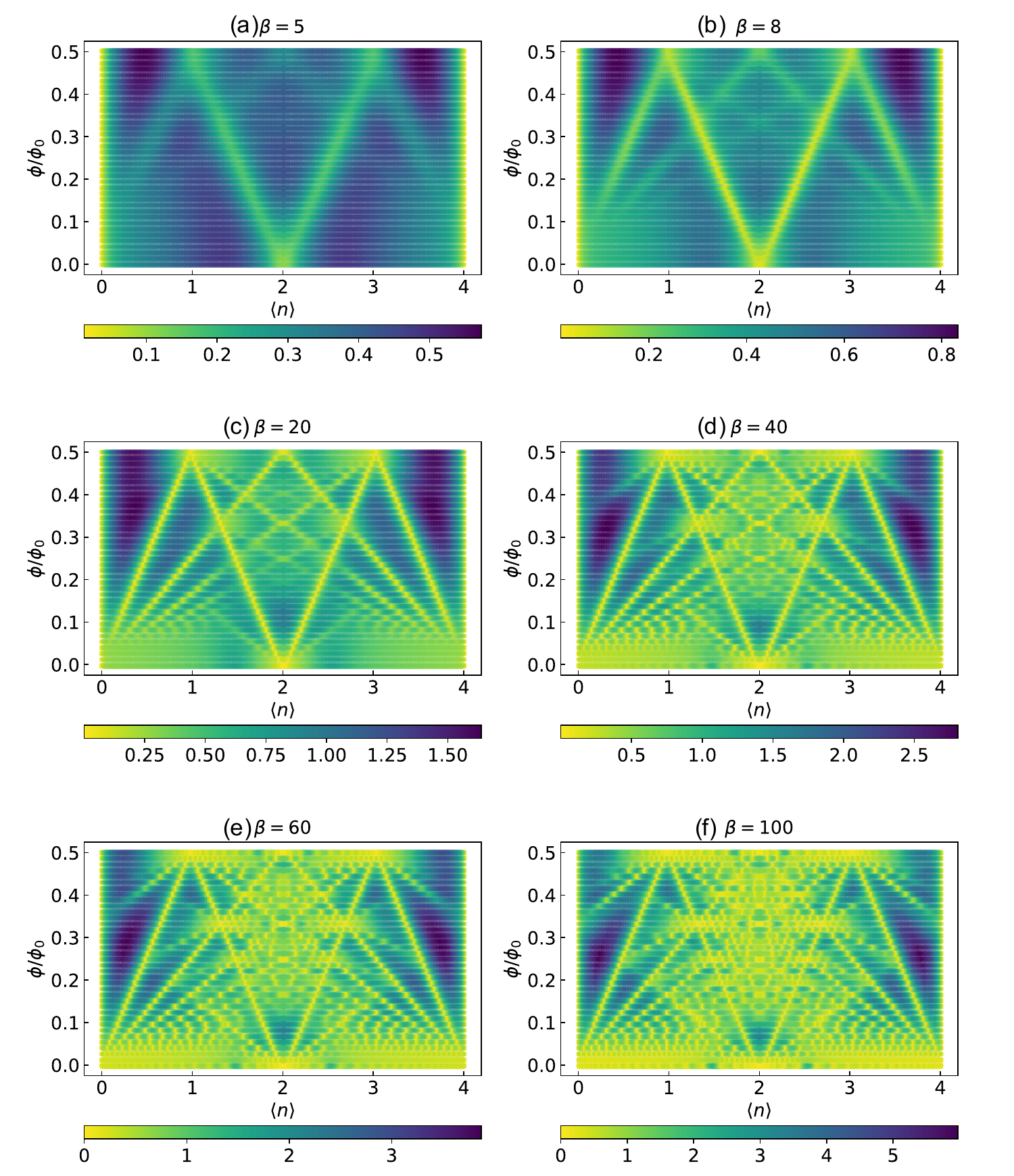}
    \caption{Non-interacting compressibility as a function of magnetic flux and density at different temperatures (a) $\beta=5/t$, (b) $\beta=8/t$, (c) $\beta=20/t$, (d) $\beta=40/t$, (e) $\beta=60/t$, (f) $\beta=100/t$. }
    \label{fig:varyT}
\end{figure*}

We show the non-interacting compressibility at different temperatures in \figdisp{fig:varyT}. As the temperature decreases, we observe more straight lines (incompressible states) representing different quantum Hall states. Particularly, for $\beta\gtrsim40/t$, incompressible Hofstadter states in the top left and right regions represented by straight lines from quarter fillings $\langle n \rangle=1$ and $\langle n \rangle=3$ at $\phi/\phi_0=0.5$. They correspond to straight lines passing through quarter-filling at $\phi/\phi_0$ and appear at $\phi/\phi_0\geq1/3$. These two features do not match the additional valleys we observe in Fig.~2(e) and Fig.~3 of the main text. Therefore, the new valleys in Fig.~2(e) and Fig.~3 of the main text are driven by the interactions.

\clearpage

\section{Supplementary Note 4: Static spin and charge susceptibility}

\begin{figure*}[ht]
    \centering
    \includegraphics[width=0.8\textwidth]{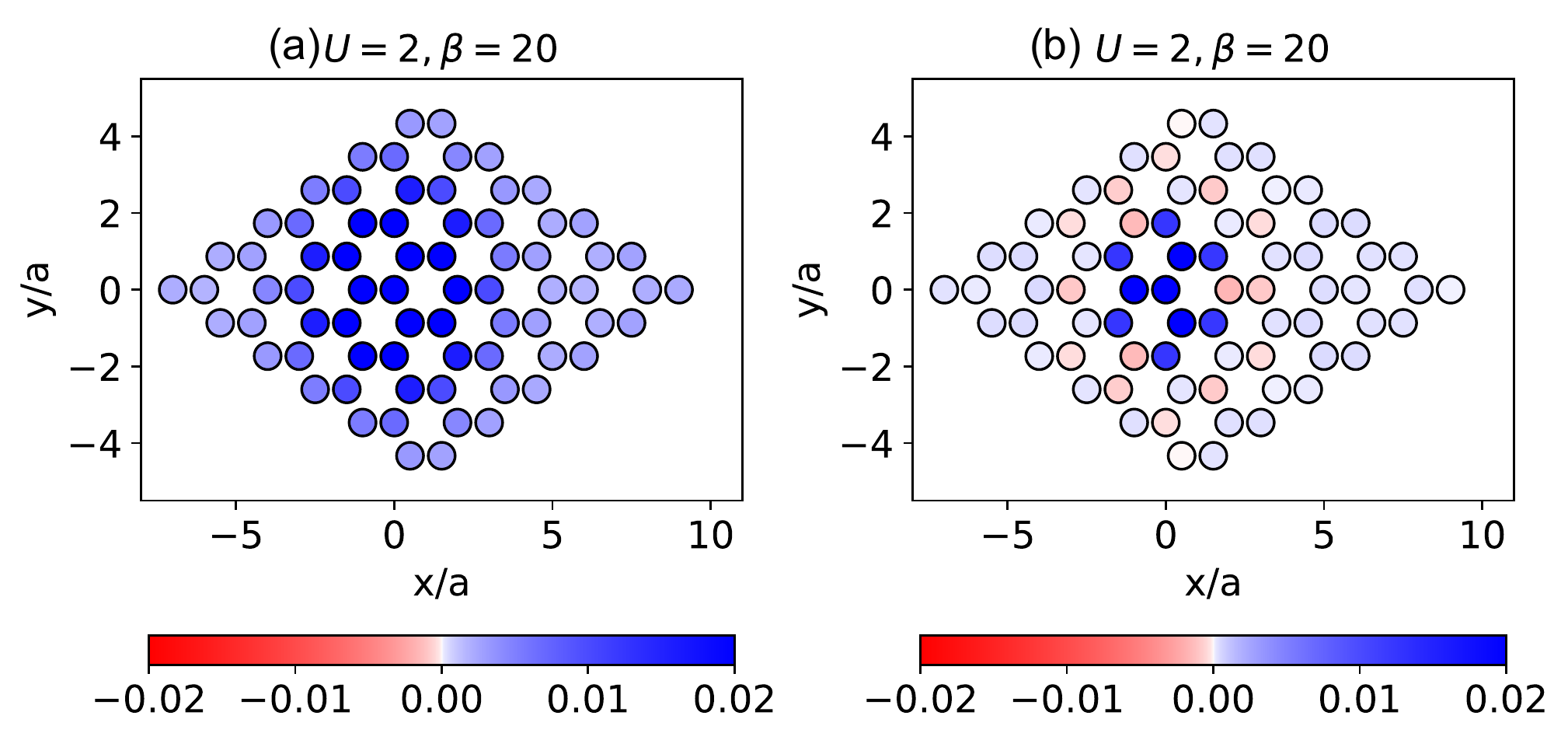}
    \caption{Zero-frequency spin (left) and charge (right) correlation at $\phi/\phi_0=11/36$ and $\langle n\rangle=11/36$ for $U/t=0$ in the first row and $U/t=2$ in the second row. The inverse temperature is $\beta=20/t$ for all panels.}
    \label{fig:sc}
\end{figure*}
The zero-frequency (static) correlation function is more sensitive for the detection of fluctuating stripe patterns at finite temperature due to the integration over imaginary time. The $\omega=0$ correlation is  defined as 
\beq
S({\bf r},\omega=0)=\frac{1}{N}\int_0^{\beta}\sum_{{\bf i}}\langle S^z_{{\bf i}+{\bf r}}(\tau) S^z_{\bf i}(0) \rangle d\tau.\label{spin}
\eeq 
for the spin case, and 
\beq
C({\bf r},\omega=0) =\frac{1}{N} \int_0^\beta \sum_{{\bf i}} \left[ \langle n_{{\bf i}+{\bf r}}(\tau) n_{j}(0)\rangle - \langle n_{{\bf i}+{\bf r}}(\tau)\rangle\langle n_{{\bf i}}(0)\rangle \right]d\tau. \label{charge}
\eeq 
for the charge correlation.  They are plotted in \figdisp{fig:sc} at $\phi/\phi_0=11/36$, $\langle n\rangle=11/36$ and $\beta=20/t$ for $U/t=2$. Turning on the interaction, as shown in the Fig.~3 of the main text, a sharper peak develops in the spin correlation and a dip emerges in the charge correlation for this parameter set. Correspondingly, the real-space resolved spin correlation displays a ferromagnetic pattern in \figdisp{fig:sc} (a) and charge correlation represents a two-dimensional stripe pattern in \figdisp{fig:sc} (b) signaled by a blue area in the middle surrounded by a red ring region.  The nature of the charge stripes deserves further study.

\clearpage
\section{Supplementary Note 5: The insulating gap at $\langle n\rangle=2$}

\begin{figure*}[ht]
    \centering
    \includegraphics[width=1.0\textwidth]{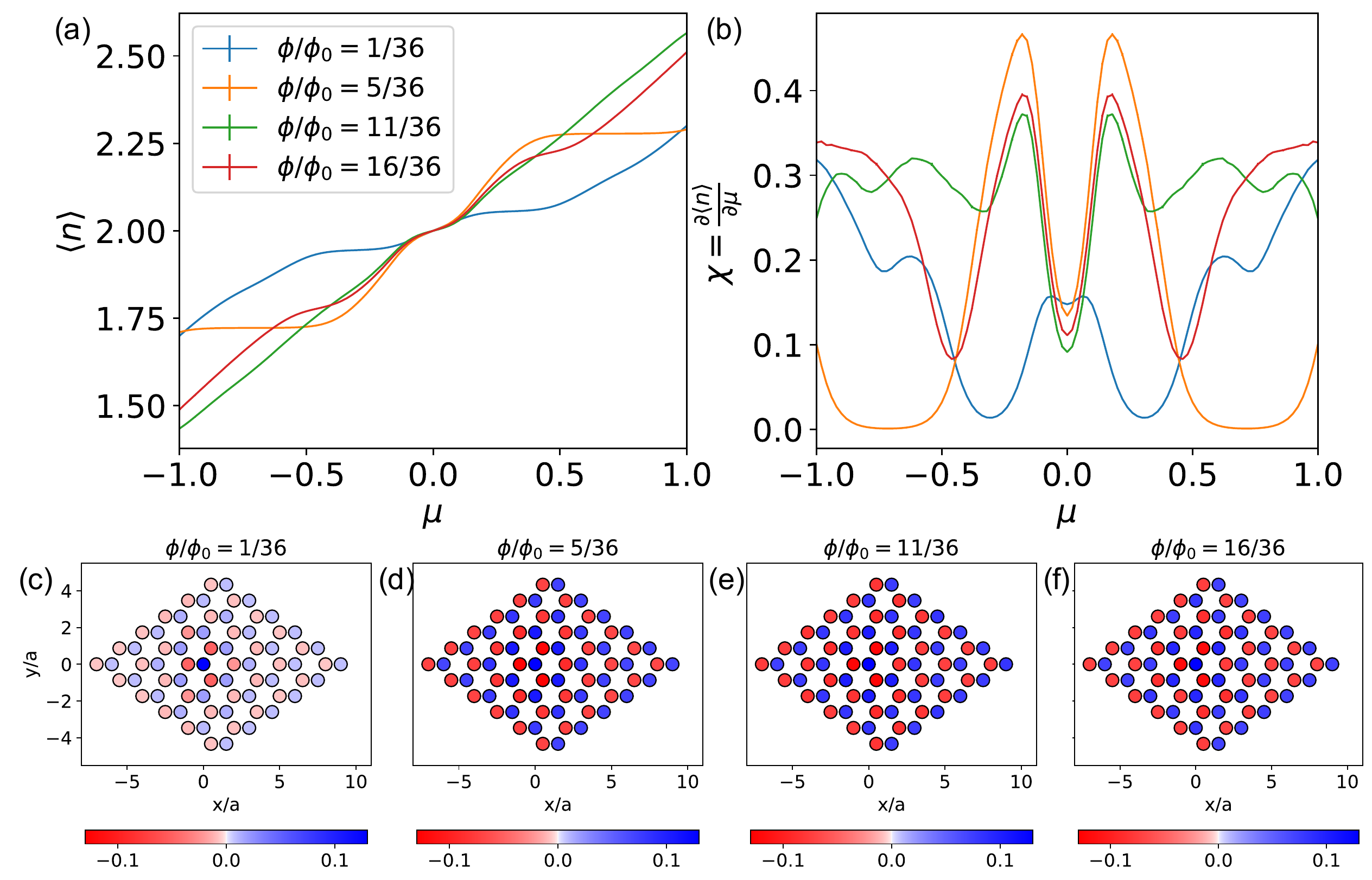}
    \caption{(a) Density and (b) compressibility as a function of chemical potential around $\langle n\rangle=2$ ($\mu=0$) at different values of magnetic fluxes. (c-f) The real-space static spin susceptibility at $\mu=0$ for each flux. The other parameters are $U/t=2$ and $\beta=20/t$. }
    \label{fig:midgapU2}
\end{figure*}

In this section, we explore the $U=2/t$ system at half-filling $\langle n\rangle=2$. When there is no external magnetic field, the system remains in the semimetal phase\cite{Assaad2,Ostmeyer} for this value of $U$. When the magnetic field reaches a certain value, \figdisp{fig:midgapU2}(a) shows that it has the tendency to form a short plateau signaling a small gap. The compressibility in \figdisp{fig:midgapU2}(b) also indicates the suppression of density of state around $\mu=0$. Specifically for $\phi/\phi_0=16/36$ (red line), the suppression is as much as that for the non-interacting quantum Hall states (Landau level with a Chern number $C=\pm4$) around $\mu=\pm0.5$. In \figdisp{fig:midgapU2}(c-f), the antiferromagnetic pattern is well established for all cases, although relatively weaker for $\phi/\phi_0=1/36$. The formation of antiferromagnetism before gap opening indicates that the insulating gap is likely to be of the Slater type.

\clearpage
\section{Supplementary Note 6: Density of states}

\begin{figure*}[ht]
    \centering
    \includegraphics[width=1.0\textwidth]{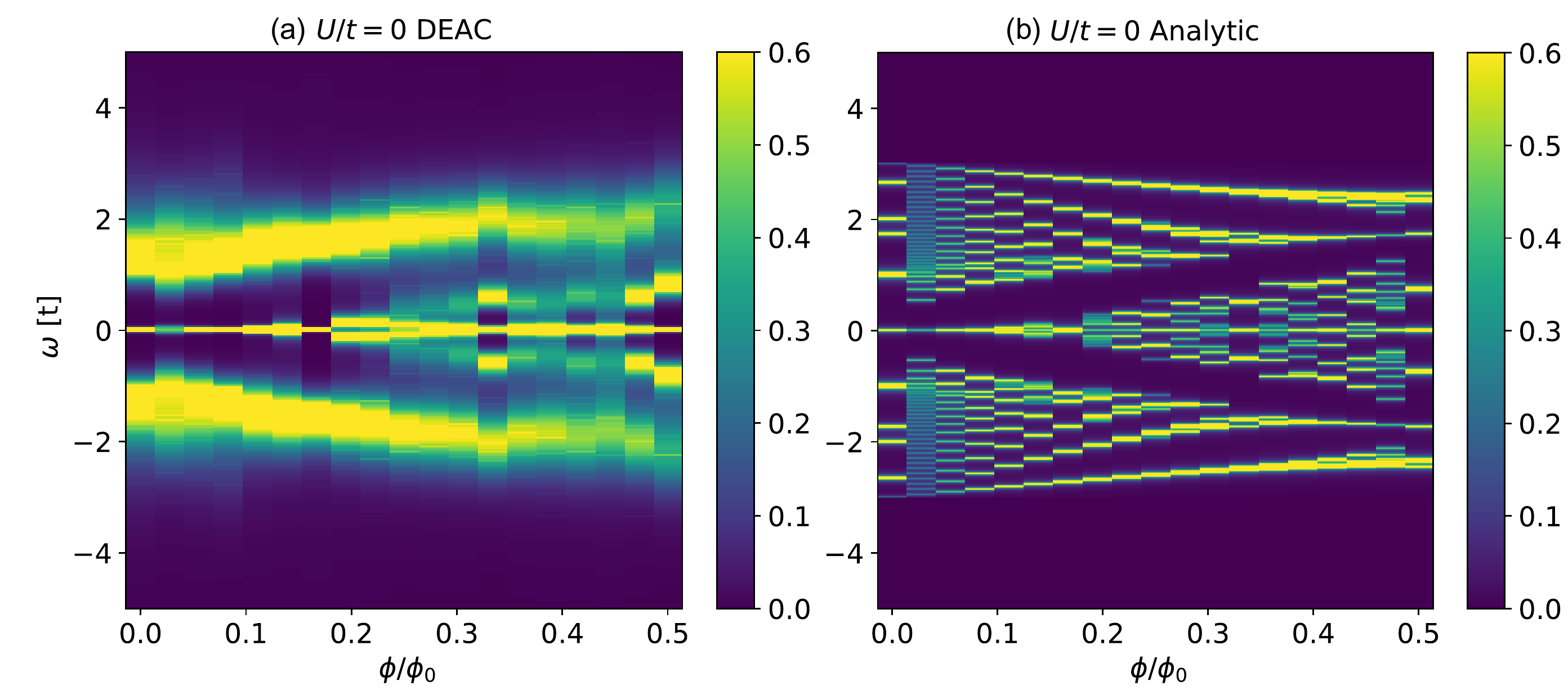}
    \caption{The non-interacting density of states from (a) an analytical-continuation of DQMC local Green functions at $\beta=30/t$ by DEAC and (b) an analytical calculation at zero temperature.}
    \label{fig:DOSnonint}
\end{figure*}

We show the comparison between the DEAC density of state at $\beta=30/t$ and the analytic density of states at zero temperature in \figdisp{fig:DOSnonint}. Note that the analytic density of states is independent of temperature, while DEAC requires lower temperature (larger $\beta$) to improve the resolution. From the comparison, we can tell that DEAC captures the main feature of the analytic result (the primary branches and gaps) although it fails to reproduce the detailed fractal pattern. 

\bibliography{reference}